\documentclass[12pt]{article}
\usepackage{epsf}
\title{ {\bf The Lorentz and CPT violating effects on the charged Higgs
boson decays $H^+\rightarrow W^+\, H^0 (h^0,A^0)$}}

\author{\vspace{1cm}\\
        {\bf E. O. Iltan}
        \thanks{E-mail address:
        eiltan@heraklit.physics.metu.edu.tr}
 \\
        Physics Department, Middle East Technical University \\
        Ankara, Turkey\\}
\date{}
\begin{document}
\setlength{\baselineskip}{24pt}
\maketitle
\setlength{\baselineskip}{7mm}
\begin{abstract}
We study the decay widths of the processes $H^+\rightarrow W^+\,
H^0\, (h^0,A^0)$, including the Lorentz violating effects and
analyze the possible CPT violating asymmetry arising from CPT odd
coefficients. We observe that these effects are too small to be
detected, since the corresponding coefficients are highly
suppressed at the low energy scale.
\end{abstract}
\thispagestyle{empty}
\newpage
\setcounter{page}{1}
\section{Introduction}
The discovery of charged Higgs boson is an effective signal about
the existence of the multi Higgs doublet structure which is lying
beyond the standard model (SM), such as two Higgs doublet model
(2HDM), minimal extension of the standard model (MSSM). There is
an extensive work on the charged Higgs boson and its possible
decays in the literature.

The charged Higgs production has been studied in several
theoretical and experimental works \cite{Diaz, L3, Andriga,
Delphison}. A search for pair-produced charged Higgs bosons was
analyzed with the L3 detector at LEP and its mass was obtained as
$m_{H^+}>76.5\, (GeV)$ \cite{L3}. The CDF and D0 collaborations
have studied $H^+$ bosons at tevatron, in the case of
$p\bar{p}\rightarrow t \bar{t}$, with at least one of the top
quark decaying via $t\rightarrow H^+ b$ and they present the
charged Higgs mass lower limits as $m_{H^+}> 77.4\, (GeV)$
\cite{Andriga}. In the recent work \cite{Delphison}, a search for
pair produced charged Higgs boson is performed using the data from
the DELPHI detector at LEP II and the existence of this particle
with mass lower than $76.7\, (74.4)GeV$ in the type I (II) 2HDM is
excluded.

The charged Higgs boson decays received a considerable interest.
In \cite{Christove, Barger} it was shown that the dominate decay
modes of the charged Higgs boson were $H^+\rightarrow \tau^+ \nu$
and $H^+\rightarrow t \bar{b}$. $H^+\rightarrow t \bar{b}$ process
in the minimal supersymmetric model (MMSM) has been analyzed in
\cite{Marcela}. Another process which was a candidate for large
branching ratio (BR) is $H^+\rightarrow W^+ h^0$ decay and it was
examined in \cite{Santos}. \cite{Diaz2} is devoted to the analysis
of $H^+\rightarrow W^+ \gamma$, $H^+\rightarrow W^+ Z$ and
$H^+\rightarrow W^+ h^0$ decays in the framework of the effective
lagrangian extension of the 2HDM. In this work the BRs have been
obtained at the order of magnitude of $10^{-5}, 10^{-1}$ and
$O(1)$, respectively. $H^+\rightarrow W^+ h^0$ decay has been
studied in MSSM in \cite{Sheng}. \cite{Akeroyd} is devoted to the
analysis of $H^+\rightarrow W^+ A^0$ decay in the framework of the
2HDM and, in this work, it was concluded that this channel might
be dominant one over a wide range of parameter space relevant at
present and future colliders. The decays of the charged Higgs
boson, including the radiative modes into decays $W^+ \gamma$ and
$W^+ Z$, has been studied mostly in the framework of the 2HDM and
MSSM in \cite{Diaz3}.

In the present work, we study the Lorentz and CPT violating
effects on the decay width ($\Gamma$) of the processes
$H^+\rightarrow H^0\, W^+ \,(h^0 \,W^+, A^0\, W^+)$ where $H^0$,
$h^0$ ($A^0$) are the scalar (pseudo scalar) Higgs fields in the
model III version of the 2HDM. A Lorentz invariant underlying
fundamental theory brings new Lorentz and CPT violating
interactions in the model III with the possible spontaneous
breaking mechanism. The extended theories like the string theory
\cite{Kos2}, the non-commutative theories \cite{Carroll}, exist at
higher scales where the Lorentz and CPT symmetries are broken
\cite{Kos1}. The space-time-varying scalar couplings can also lead
to Lorentz violating effects described by the SM extensions
\cite{Perry}. The Lorentz violation ensures tiny interactions at
the low energy level and they are presented in the SM extension in
\cite{Colladay, Lehnert}.

In the literature there are various studies on the general Lorentz
and CPT violating effects in the framework of Quantum Electro
Dynamics (QED),  the noncommutative space time, the Wess-Zumino
model and on the restriction of the Lorentz violating coefficients
coming from the experiments \cite{Kos4,Russell}. Furthermore, some
phenomenological works have been done on the the Lorentz and CPT
violating effects in the SM and the model III extensions
\cite{EiltmuegamLrVio} and it was observed that these tiny effects
are far from the detection in the experiments. \cite{Turan} is
devoted to the bounds on the CPT-even asymmetric (symmetric )
coefficients which arise from the one-loop contributions to the
photon propagator  (from the equivalent $c_{\mu\nu}$ coefficients
in the fermion sector), and those from the CPT-odd coefficient
which arise from bounds on the vacuum expectation value of the
Z-boson.

The present work is devoted to the prediction of Lorentz and CPT
violating effects on the decay widths of the charged Higgs $H^+$
decays into $W^+$ boson and  scalar (pseudo scalar) Higgs bosons,
in the model III version of 2HDM.  We study the relative behavior
of the Lorentz violating effects which are carried by tiny CPT
even and CPT odd coefficients and estimate the possible CPT
violating asymmetry arising from CPT odd one.  We observe that
these additional effects are too weak to be detected in the
present experiments, since the coefficients driving those effects
lie in the range which is regulated by the suppression scale taken
as the ratio of the light one, of the order of the electroweak
scale, to the one of the order of the Planck mass \cite{Russell}.

The paper is organized as follows: In Section 2, we present the
theoretical expression of the decay width $\Gamma$,  for the
$H^+\rightarrow H^0\, W^+  \,(h^0 \,W^+, A^0\, W^+)$ decays, with
the inclusion of the Lorentz and CPT violating effects. Section 3
is devoted to discussion and our conclusions.
\section{The Lorentz and CPT violating effects on charged Higgs decays into
$W^+\, S$ ($S=H^0, h^0, A^0$).} This section is devoted to the
Lorentz and CPT violating effects on the $\Gamma$ of the charged
Higgs decays, $H^+\rightarrow H^0\, W^+  \,(h^0 \,W^+, A^0\, W^+)$
where $H^0$, $h^0$ ($A^0$) are the scalar, (pseudo scalar) Higgs
fields in the 2HDM (see \cite{Haber} for review)). The charged
Higgs decays under consideration exist in the tree level and the
Lorentz violating effects appear with the addition new
interactions which may come from more fundamental theory in the
Planck scale. These tiny effects are regulated by the new
coefficients, having small numerical values which has the
suppression scale proportional to the ratio of the mass in the
electroweak scale to the one in the Planck scale.

The charged Higgs decays, $H^+\rightarrow H^0\, W^+  \,(h^0 \,W^+,
A^0\, W^+)$ are induced by the so called kinetic part of the
lagrangian and, in the 2HDM, it reads
\begin{eqnarray}
{\cal{L}}_{Higgs}&=&(D_\mu \phi_1)^\dagger D_\mu \phi_1 +(D_\mu
\phi_2)^\dagger D_\mu \phi_2\, , \label{lagrangianHiggs}
\end{eqnarray}
where $\phi_{1,2}$ are the Higgs scalar doublets in a suitable
basis (see \cite{Atwood} for example)
\begin{eqnarray}
\phi_{1}=\frac{1}{\sqrt{2}}\left[\left(\begin{array}{c c}
0\\v+\bar{H}^{0}\end{array}\right)\; + \left(\begin{array}{c c}
\sqrt{2} \chi^{+}\\ i \chi^{0}\end{array}\right) \right]\, ;
\phi_{2}=\frac{1}{\sqrt{2}}\left(\begin{array}{c c} \sqrt{2}
H^{+}\\ H_1+i H_2 \end{array}\right) \,\, , \label{choice}
\end{eqnarray}
with the vacuum expectation values
\begin{eqnarray}
<\phi_{1}>=\frac{1}{\sqrt{2}}\left(\begin{array}{c c}
0\\v\end{array}\right) \,  \, ; <\phi_{2}>=0 \,\, .
\label{choice2}
\end{eqnarray}

Here the neutral bosons $\bar{H}_0$, $H_1$ and $H_2$ are defined
in terms of the mass eigenstates $H_0$ ,$h_0$ and $A_0$ as
\begin{eqnarray}
\bar{H}_{0}&=&(H_{0} cos \alpha - h_0 sin\alpha) \nonumber \, ,\\
H_{1}&=&( h_{0} cos \alpha + H_0 sin\alpha) \nonumber \, ,\\
H_{2}&=&A_0 \,\,, \label{neutrbos}
\end{eqnarray}
where $\alpha$ is the mixing angle and $v$ is proportional to the
vacuum expectation value of the doublet $\phi_1$ (eq.
(\ref{choice2})). In eq. (\ref{lagrangianHiggs}) $D_{\mu}$ is the
covariant derivative, $D_{\mu}=\partial_{\mu}+\frac{i g}{2}
\tau.W_{\mu}+\frac{i g'}{2} Y B_{\mu}$, $\tau$ is the Pauli spin
matrix, Y is the weak hypercharge, $B_{\mu}$ ($W_{\mu}$) is the
$U(1)_Y$ ($SU(2)_L$ triplet) gauge field.

The additional part due to the Lorentz violating effects can be
represented by the CPT-even and CPT-odd lagrangian \cite{Colladay}
as
\begin{eqnarray}
{\cal{L}}^{CPT-even}_{Higgs\, LorVio}&=&\frac{1}{2}\,
(k_{\phi\phi})^{\mu\nu} \,\Bigg( (D_\mu \phi_1)^\dagger D_\nu
\phi_1+(D_\mu \phi_2)^\dagger D_\nu \phi_2 \Bigg)+h.c. \nonumber
\\
&-& \frac{1}{2}\,(k_{\phi W})^{\mu\nu}\,
(\phi^\dagger_1\,W_{\mu\nu}\,\phi_1+\phi^\dagger_2\,W_{\mu\nu}\,\phi_2)
\, , \nonumber \\
{\cal{L}}^{CPT-odd}_{Higgs\, LorVio}&=&i (k_{\phi})^{\mu}\,
(\phi^\dagger_1\,D_{\mu}\,\phi_1+\phi^\dagger_2\,D_{\mu}\,\phi_2)+h.c\,
, \label{lagrangianHiggsLorVio}
\end{eqnarray}
where the coefficients $k_{\phi\phi}$ ($k_{\phi W}$) are
dimensionless with symmetric real and antisymmetric imaginary
parts (has dimension of mass and real antisymmetric). The CPT odd
coefficient $k_{\phi}$ is a complex number and has dimension of
mass. Here $W_{\mu\nu}$ is the field tensor which is defined in
terms of the gauge field $W_{\mu}$,
\begin{eqnarray}
W_{\mu\nu}&=& \partial_{\mu} W_{\nu}-\partial_{\nu} W_{\mu}+i g
[W_{\mu},W_{\nu}]\, . \label{WBmunu}
\end{eqnarray}

This Lagrangian brings new Lorentz violating corrections driven by
the vertices
\begin{eqnarray}
(V^{even})^\mu&=&\frac{1}{2}\,g\,(k^{Sym}_{\phi
\phi})^{\mu\nu}\,(k-p)_\nu+ i\,(k_{\phi W})^{\mu\nu}\,q_\nu\,,
\nonumber \\
(V^{odd})^\mu&=&\frac{1}{2}\,g\,(k_{\phi}^{\mu}+k_{\phi}^{\mu
+})\, . \label{vertHplWplS}
\end{eqnarray}
where p (k; q) is the four momentum vector of incoming $H^{+}\,
(S=H^0,h^0,A^0; W^+)$, $sin\alpha \,(V^{even})^\mu$ ($cos\alpha
\,(V^{even})^\mu$, $-i\,(V^{even})^\mu$) is the CPT-even Lorentz
violating vertex and $sin\alpha \,V^{\mu\,odd}$ ($cos\alpha
\,V^{\mu\,odd}$, $(-i)\,V^{\mu\,odd}$) is the CPT-odd Lorentz
violating vertex for the $H^+\rightarrow H^0\, (h^0, A^0)\,W^+$
decay.

It is  well known that the invariant phase-space elements in the
presence of Lorentz violation are modified \cite{Potting}. In the
case that there are no Lorentz violating effects, the expression
for decay width in the $H^+$ boson rest frame reads
\begin{eqnarray}
d\Gamma&=&\frac{(2\pi)^4}{2\,m_{H^+}}\,
\delta^{(4)}(p_{H^+}-q_W-q_S)\, \frac{d^3
q_W}{(2\pi)^3\,2\,E_W}\,\frac{d^3 q_S}{(2\pi)^3\,2\,E_S}
 \nonumber \\ &\times& |M|^2 (p_{H^+},q_W,q_S)
\label{DecWid}
\end{eqnarray}
with the four momentum vector of $H^+$ boson ($W^+, S$) $p_{H^+}$
($q_W, \, q_S$), and the matrix element $M$ for the process
$H^{+}\rightarrow S\, W^+$.  With the inclusion of the new Lorentz
violating parameters in the neutral Higgs sector, the $S$ boson
dispersion relation changes and this induces an additional part in
the phase space element $\frac{d^3 q_S}{(2\pi)^3\,2\,E_S}$.

The variational procedure generates the equation (see eqs.
(\ref{lagrangianHiggs}) and (\ref{lagrangianHiggsLorVio}))
\begin{equation}
(-\partial^2 - m_S^2-Re[k_{\phi\phi}^{\mu \nu}] \,
\partial_{\mu}\, \partial_{\nu}- 2 \,Im[k_\phi^\mu]\, \partial_\mu)\, \, S=0 \,
,
\end{equation}
which leads to the dispersion relation
\begin{equation}
\Big(q_S^2\,(1+|k_{\phi\phi}^{Sym}|)-m_S^2 \Big)^2+4\,
\Big(Im[k_\phi^\mu]\,q_{S\,\mu}\Big)^2=0 \nonumber \, .
\end{equation}
Here take the special parametrization where the symmetric part of
the coefficient $k_{\phi\phi}^{\mu\nu}$ is proportional to the
identity:
\begin{eqnarray}
k_{\phi\phi}^{\mu\nu}&=& \delta^{\mu\nu}\, |k_{\phi\phi}^{Sym}|+
k_{\phi\phi}^{ASym\, \mu\nu} \, . \label{kpar}
\end{eqnarray}
and, in the following, we assume that $k_\phi^\mu$ is real.
Finally the dispersion relation becomes
\begin{equation}
q_S^2\,(1+|k_{\phi\phi}^{Sym}|)-m_S^2=0 \nonumber \, ,
\end{equation}
and the energy eigenvalues are obtained as
\begin{eqnarray}
E_S^{\pm}=\pm \sqrt{\frac{m_S^2+(1+|k_{\phi\phi}^{Sym}|)\,
\vec{q}_S^{\,\,2}} {(1+|k_{\phi\phi}^{Sym}|)}} \nonumber \, .
\end{eqnarray}
Using the vertices presented eq. (\ref{vertHplWplS}) and the
modified phase space element, the decay width $\Gamma$ in the
$H^\pm$ boson rest frame, linear in the Lorentz violating
parameters, is obtained as
\begin{eqnarray}
\Gamma^S=\frac{1}{64\,\pi\, m^3_{H^+}}\, g^2\, \sqrt{\Delta}\,
(\Gamma^S_0+\Gamma^S_{LorVio}) \label{vertH0ff}
\end{eqnarray}
where
\begin{eqnarray}
\Gamma^S_0&=&\frac{1}{m^2_W}\, f_1^S \xi \nonumber \, , \\
\Gamma^S_{LorVio}&=& \frac{1}{m^2_W\,m_{H^+}}\,\xi\, \Bigg( -2\,
Re[k_{\phi}^0]\,f_1^S + \frac{1}{m_{H^+}}\,k^{Sym}_{\phi
\phi}\,\,f_2^S
\,(1+\frac{m_S^2\,m_{H^+}^2}{\Delta})\bigg)\label{Gamma0LorVio}
\end{eqnarray}
with
\begin{eqnarray}
\Delta&=&(m_S^2+m_{H^+}^2-m_W^2)^2-4\,m_S^2\,m_{H^+}^2\, ,
\nonumber \\
 f_1^S&=&\beta \Bigg(
m^4_{S}+(m^2_{H^+}-m_W^2)^2-2\,m^2_{S}\,(m^2_{H^+}+m_W^2) \Bigg)
\, ,
\nonumber \\
f_2^S&=&\beta\,
\Bigg(-m^6_{S}+3\,m^4_{S}\,(m^2_{H^+}+m_W^2)+(m^2_{H^+}-m_W^2)^2\,
(m^2_{H^+}+m_W^2) \nonumber
\\ &-&
m^2_{S}\,(3\,m^4_{H^+}+2\,m^2_{H^+}\,m_W^2+3\,m_W^4) \Bigg)
\label{f1f2}
\end{eqnarray}
Here the parameters $\xi$ and $\beta$ read $\xi=sin^2\alpha\,
(cos^2\alpha, 1)$ for $S=H^0 \,(h^0, A^0)$ and $\beta=1\,(-1)$ for
$S=H^0,h^0 \,(A^0)$. Notice that the additional term
$\frac{m_S^2\,m_{H^+}^2}{\Delta}$ in the parenthesis, in eq.
(\ref{Gamma0LorVio}), is due to the modified phase factor.

Eq. (\ref{Gamma0LorVio}) shows that the Lorentz violating effects
enter into decay width of charged Higgs boson with the CPT even
$k_{\phi\phi}^{Sym}$ and CPT odd $k_{\phi}$ coefficients. The
latter one is responsible for the tiny CPT asymmetry in these
decays, namely
\begin{eqnarray}
A_{CPT}=\frac{\Gamma-\bar{\Gamma}}{\Gamma+\bar{\Gamma}}\, ,
\label{ACPT}
\end{eqnarray}
where $\bar{\Gamma}$ the CPT conjugate of the $\Gamma$.
\section{Discussion}
The SM model is invariant under Lorentz and CPT transformations.
However, the spontaneous Lorentz violation in the Lorentz and CPT
invariant more  fundamental theory at the Planck scale brings new
interactions in the lower level where the ordinary SM lies. These
additional interactions are naturally suppressed and their
strengths are proportional to the ratio of the light mass at the
order of $m_{f,W,Z}$ to the one of the order of the Planck mass.
This leads to a range $10^{-23}-10^{-17}$ \cite{Russell} for the
coefficients which carry the Lorentz and CPT violating effects.
Notice that the first (second) number represent the electron mass
$m_e$ ($m_{EW}\sim 250\,GeV$) scale.

In this section, we study the Lorentz and CPT violating effects on
the decay widths of the charged Higgs $H^+$ decays $H^+\rightarrow
H^0 (h^0,A^0)$ and predict a possible CPT violating asymmetry
arising from CPT odd coefficients. The Lorentz violation is
regulated by CPT even $k^{Sym}_{\phi \phi}$, $k_{\phi W}$ and  CPT
odd $k_{\phi}$ coefficients and the $\Gamma$ of the decays under
consideration depends on $k^{Sym}_{\phi \phi}$ and $k_{\phi}$
which leads to tiny CPT asymmetry.

Now, for completeness, we start with the analysis of the $\Gamma
(H^+\rightarrow W^+\, H^0 (h^0,A^0))$ in the model III without
Lorentz violating effects.

\begin{figure}[htb]
\vskip -3.0truein \centering \epsfxsize=6.8in
\leavevmode\epsffile{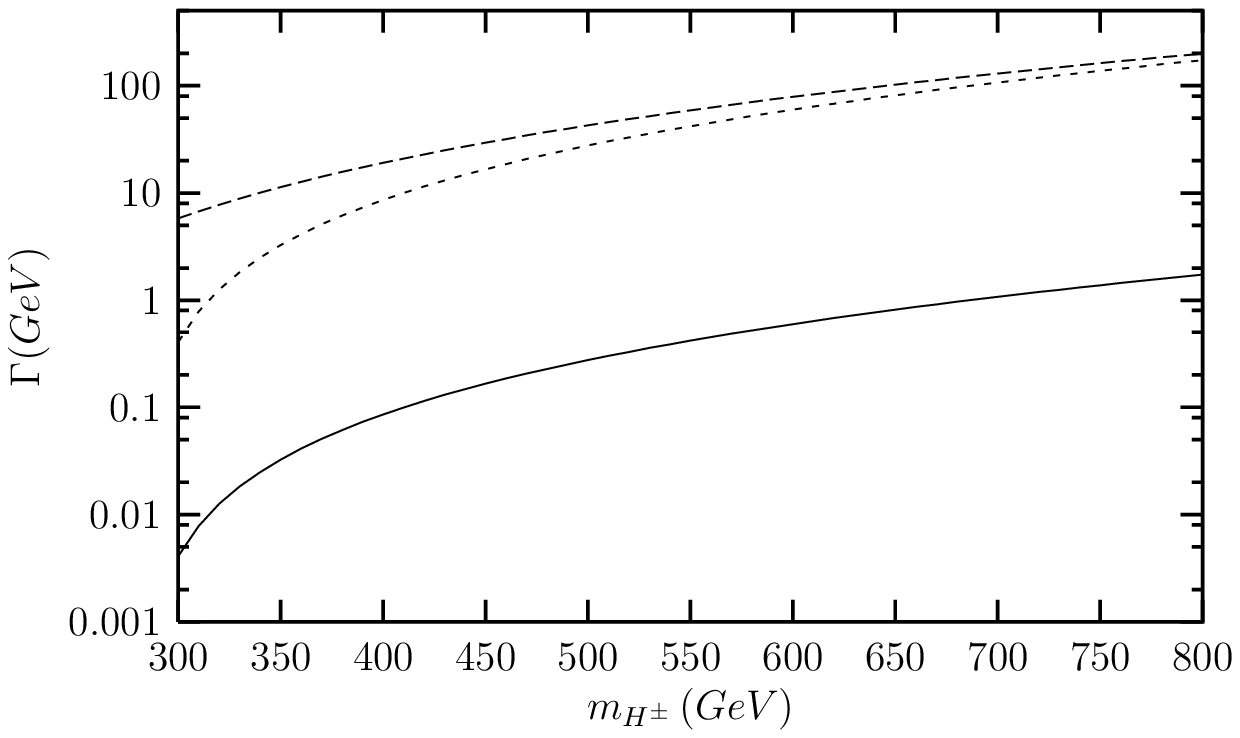} \vskip -3.0truein \caption[]{ The
$m_{H^+}$ dependence of the $\Gamma$ for $H^+\rightarrow W^+\,
H^0\, (h^0,A^0)$ decay, for the fixed values $m_{H^0}=200\,GeV,\,
m_{h^0}=100\,GeV,\,m_{A^0}=200\,GeV$, $sin\alpha=0.1$. Here solid
(dashed, small dashed) line represents the dependence for $H^0
\,(h^0,A^0)$ output.} \label{DWHpl}
\end{figure}

\begin{figure}[htb]
\vskip -3.0truein \centering \epsfxsize=6.8in
\leavevmode\epsffile{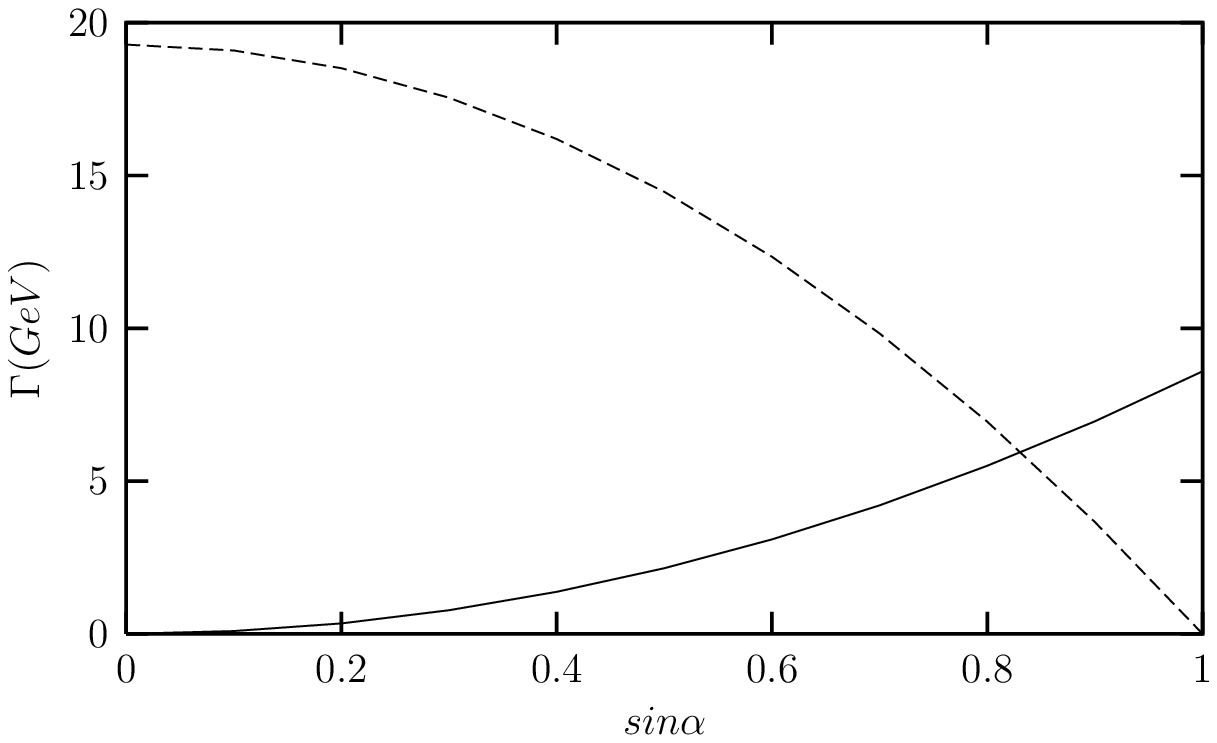} \vskip -3.0truein \caption[]{The
$sin\alpha$ dependence of the $\Gamma$ for $H^+\rightarrow W^+
\,H^0\, (h^0)$ decay, for the fixed values $m_{H^+}=400\,GeV,\,
m_{H^0}=200\,GeV,\, m_{h^0}=100\,GeV$. Here solid (dashed) line
represents the dependence for $H^0 \,(h^0)$ output.} \label{DWsin}
\end{figure}

Fig. \ref{DWHpl} (\ref{DWsin}) is devoted to the $m_{H^+}$
($sin\alpha$) dependence of the $\Gamma$ for $H^0,h^0,A^0$
($H^0,h^0$) outputs and for $m_{H^0}=200\,GeV,\,
m_{h^0}=100\,GeV,\,m_{A^0}=200\,GeV$, $sin\alpha=0.1$
($m_{H^+}=400\,GeV,\, m_{H^0}=200\,GeV,\,
m_{h^0}=100\,GeV,\,m_{A^0}=200\,GeV$). In Fig. \ref{DWHpl} solid
(dashed, small dashed) line represents the dependence for $H^0
\,(h^0,A^0)$ output. The $\Gamma$ reaches $20\,(10)\,GeV$ for
$h^0\,(A^0)$ output, for the charged Higgs mass values $\sim
400\,GeV$. This is almost two order larger compared to the case
where the output scalar is $H^0$. Notice that, here, we consider a
weak mixing between neutral Higgs bosons and this results in a
suppressed $\Gamma$ for $H^0$ output. Fig. \ref{DWsin} shows the
effect of mixing of neutral Higgs scalars on the $\Gamma$s of
$H^+\rightarrow W^+\,H^0 (h^0)$ decays.

The addition of Lorentz violating effects brings small
contributions to the $\Gamma$ and in the following we study the
relative behaviors of the new coefficients driving the Lorentz
violation.

\begin{figure}[htb]
\vskip -3.0truein \centering \epsfxsize=6.8in
\leavevmode\epsffile{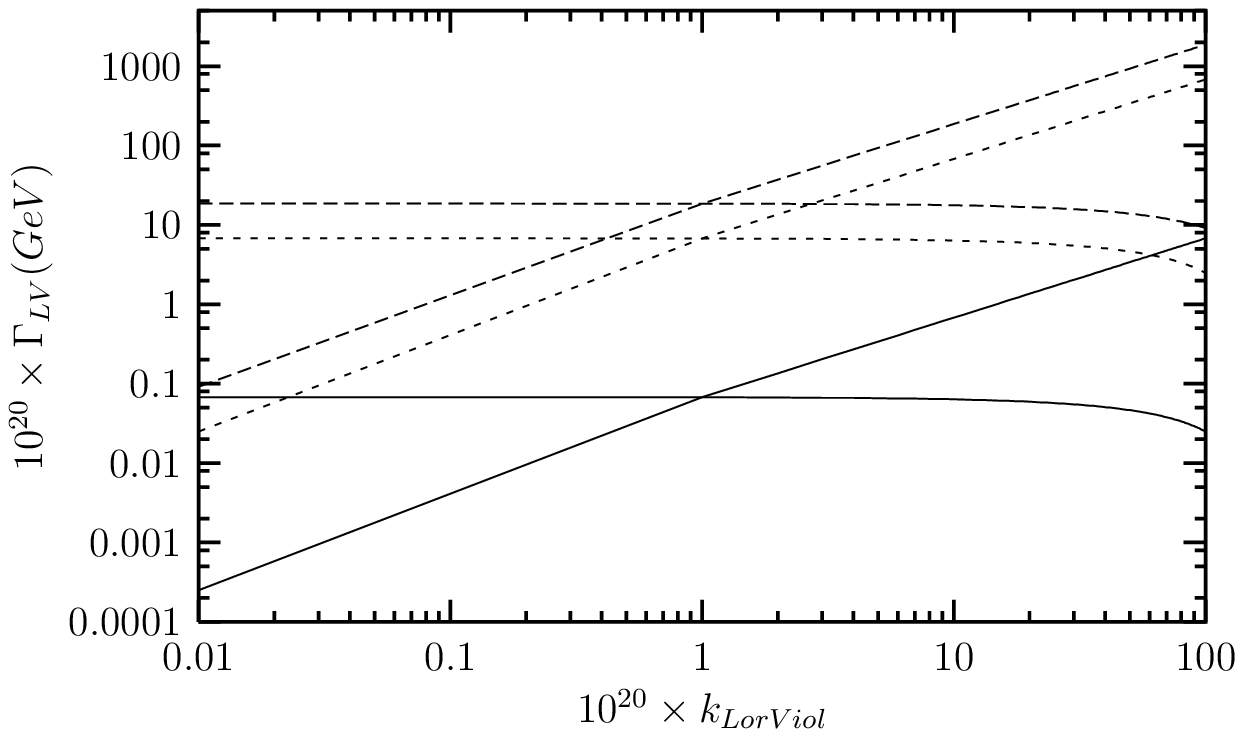} \vskip -3.0truein
\caption[]{The coefficient $k_{\phi\phi}^{Sym}\,(Re[k^0_{\phi}])$
dependence of the $\Gamma_{LV}$ for $H^+\rightarrow W^+ \,S$,
$S=\,H^0,\, h^0,\, A^0$ decays for the fixed values
$m_{H^+}=400\,GeV, m_{H^0}=200\,GeV,\,
m_{h^0}=100\,GeV,\,m_{A^0}=200\,GeV$, $sin\alpha=0.1$ and for
$Re[k^0_{\phi}]=10^{-20}\,GeV$ ($k_{\phi\phi}^{Sym}=10^{-20}$) .
Here solid, dashed, small dashed inclined (almost straight) lines
represent the dependence of $\Gamma_{LV}$ to the coefficient
$k_{\phi\phi}^{Sym}\,(Re[k^0_{\phi}])$ for $H^0,h^0,A^0$ outputs,
respectively.} \label{DWkLorVio}
\end{figure}

In Fig. \ref{DWkLorVio}, we present the coefficient
$k_{\phi\phi}^{Sym}\,(Re[k^0_{\phi}])$ dependence of magnitude of
the Lorentz violating part of the $\Gamma$, $\Gamma_{LV}$, for
$H^0,h^0,A^0$ outputs and for the fixed values, $m_{H^+}=400\,GeV,
m_{H^0}=200\,GeV,\, m_{h^0}=100\,GeV,\,m_{A^0}=200\,GeV$,
$sin\alpha=0.1$, $Re[k^0_{\phi}]=10^{-20}\, GeV$
($k_{\phi\phi}^{Sym}=10^{-20}$) . Here solid, dashed, small dashed
inclined (almost straight) lines represent the dependence of
$\Gamma_{LV}$ to the coefficient
$k_{\phi\phi}^{Sym}\,(Re[k^0_{\phi}])$ for $H^0,h^0,A^0$ outputs
respectively. The $\Gamma_{LV}$ is relatively more sensitive to
the CPT even coefficient $k_{\phi\phi}^{Sym}$ compared to the CPT
odd one $Re[k^0_{\phi}]$. The $\Gamma_{LV}$ lies in the range
$10^{-21}\, GeV\leq \Gamma \leq 10^{-17}\, GeV$ in the expected
region of $k_{\phi\phi}^{Sym}$, $10^{-22}\leq k_{\phi\phi}^{Sym}
\leq 10^{-18}$ for $h^0$ output. For $A^0$ output the upper and
lower limits of the range for $\Gamma_{LV}$ becomes almost half of
the previous one. These limits are suppressed in the case of $H^0$
output due to the weak mixing between neutral Higgs bosons.
Increasing values of the CPT odd coefficient $Re[k^0_{\phi}]$
causes slightly to decrease the $\Gamma_{LV}$ for the fixed values
of $k_{\phi\phi}^{Sym}$.

\begin{figure}[htb]
\vskip -3.0truein \centering \epsfxsize=6.8in
\leavevmode\epsffile{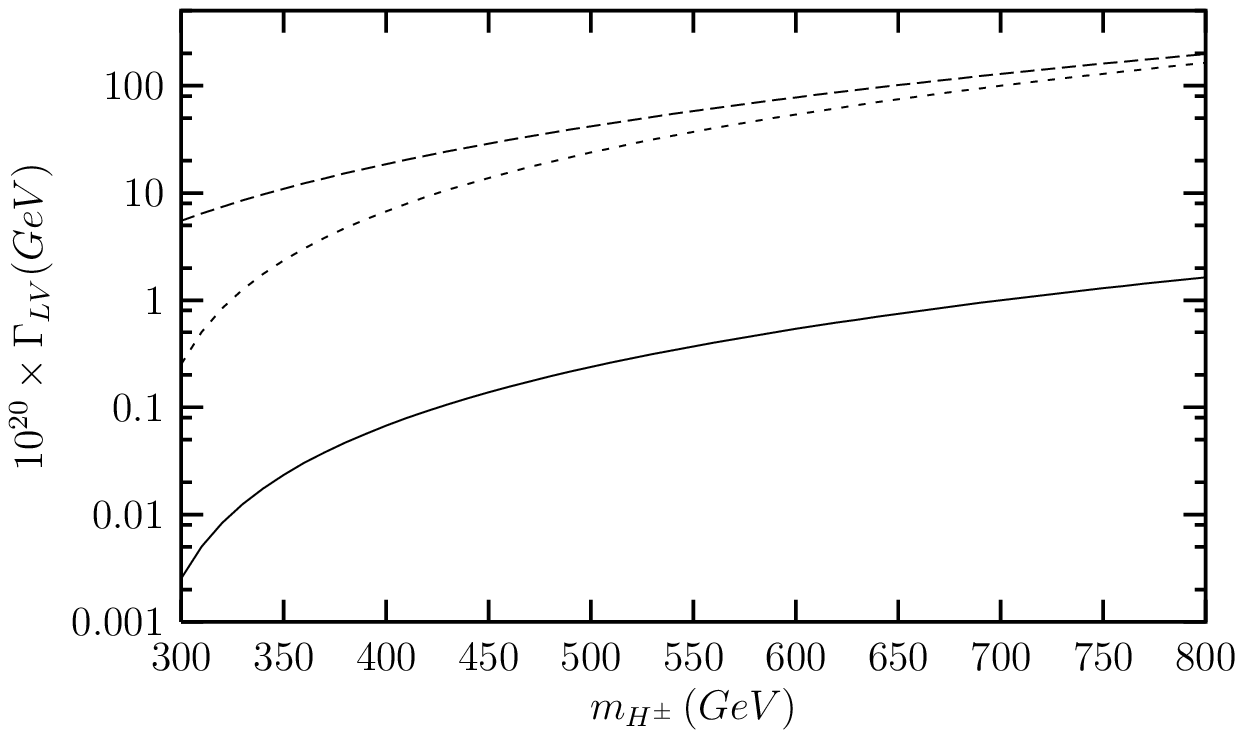} \vskip -3.0truein
\caption[]{The  Higgs mass $m_{H^+}$ dependence of the
$\Gamma_{LV}$ for $H^+\rightarrow W^+\,H^0\, (h^0,A^0)$ decay, for
$k_{\phi\phi}^{Sym}=10^{-20}$, $Re[k^0_{\phi}]=10^{-20}$, and for
$m_{H^0}=200\,GeV,\, m_{h^0}=100\,GeV,\,m_{A^0}=200\,GeV$,
$sin\alpha=0.1$. Here solid (dashed, small dashed) line represents
the dependence for $H^0 \,(h^0,A^0)$ output.} \label{DWLorVioHpl}
\end{figure}

\begin{figure}[htb]
\vskip -3.0truein \centering \epsfxsize=6.8in
\leavevmode\epsffile{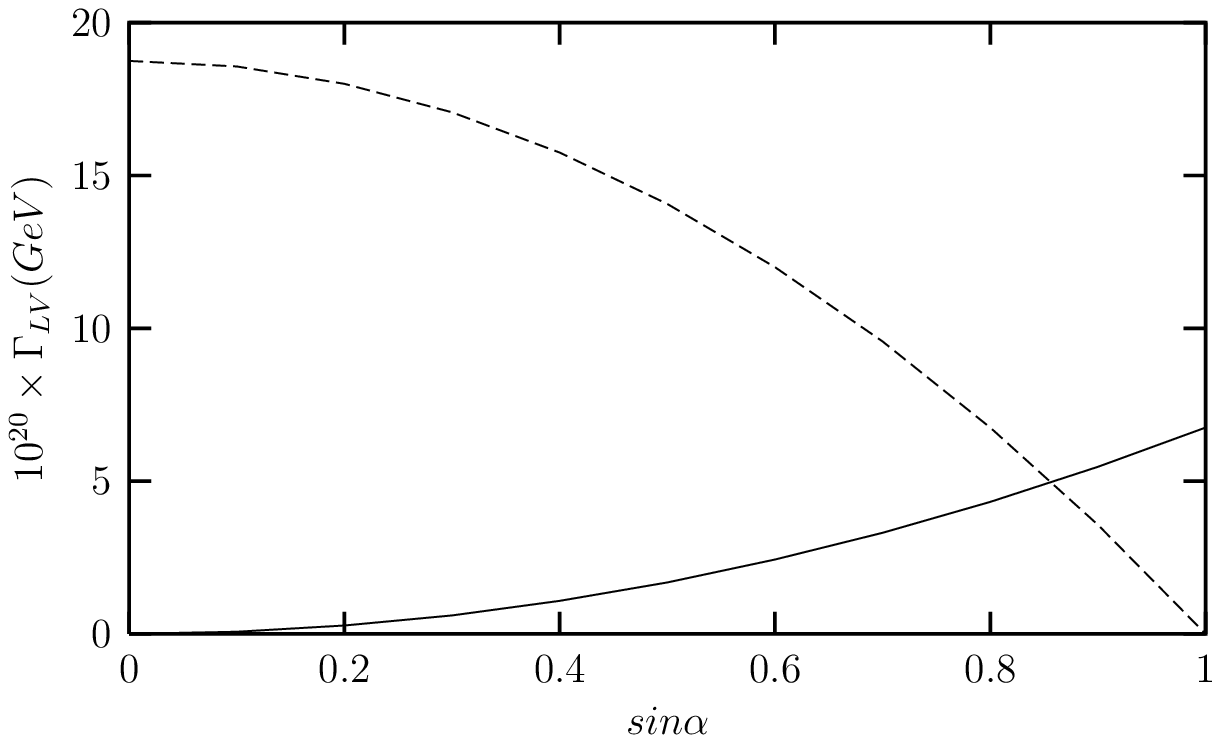} \vskip -3.0truein
\caption[]{The  $sin\alpha$ dependence of the $\Gamma_{LV}$ for
$H^+\rightarrow W^+\, H^0\, (h^0)$ decay, for
$k_{\phi\phi}^{Sym}=10^{-20}$, $Re[k^0_{\phi}]=10^{-20}\,GeV$ and
for $m_{H^+}=400\,GeV,\, m_{H^0}=200\,GeV,\, m_{h^0}=100\,GeV$.
Here solid (dashed) line represents the dependence for $H^0
\,(h^0)$ output.} \label{DWLorViosin}
\end{figure}

Fig. \ref{DWLorVioHpl} (\ref{DWLorViosin}) represents the
$m_{H^+}$ ($sin\alpha$) dependence of the $\Gamma_{LV}$ for
$k_{\phi\phi}^{Sym}=10^{-20}$, $Re[k^0_{\phi}]=10^{-20}\, GeV$,
$H^0,h^0,A^0$ ($H^0,h^0$) outputs and $m_{H^0}=200\,GeV,\,
m_{h^0}=100\,GeV,\,m_{A^0}=200\,GeV$, $sin\alpha=0.1$
($m_{H^+}=400\,GeV,\, m_{H^0}=200\,GeV,\,
m_{h^0}=100\,GeV,\,m_{A^0}=200\,GeV$). In Fig. \ref{DWLorVioHpl}
(\ref{DWLorViosin}) solid, dashed, small dashed (solid, dashed)
lines represent the $m_{H^+}$ ($sin\,\alpha$) dependence of the
$\Gamma_{LV}$ for $H^0, h^0, A^0$ ($H^0, h^0$) output. The
$\Gamma_{LV}$ reaches $20\,(10)\,10^{-20} \,GeV$ for $h^0\,(A^0)$
output, for the charged Higgs mass values $\sim 400\,GeV$. This is
almost two order larger compared to the case where the output
scalar is $H^0$, similar to the case where the SM decay width is
considered. Fig. \ref{DWLorViosin} shows the effect of mixing on
the $\Gamma_{LV}$ for neutral scalar outputs.

Finally we analyze the CPT violating asymmetry of the decays
studied (see eq. (\ref{ACPT})). This asymmetry is due to the
existence of the CPT odd parameter $k_{\phi}$ and it enters into
the $\Gamma$ as $Re[k^0_{\phi}]$.

\begin{figure}[htb]
\vskip -3.0truein \centering \epsfxsize=6.8in
\leavevmode\epsffile{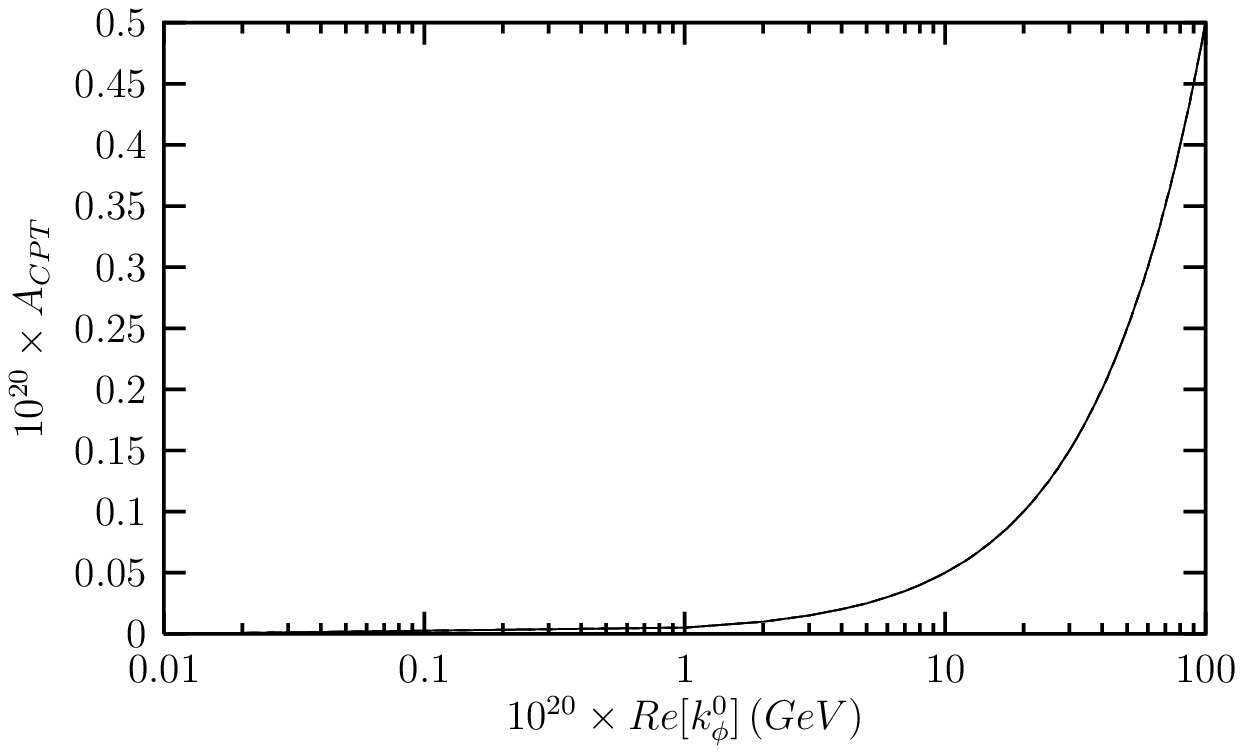} \vskip -3.0truein
\caption[]{$Re[k^0_{\phi}]$ dependence of the $A_{CPT}$ for
$H^+\rightarrow H^0\, (h^0,A^0)$ decay, for
$k_{\phi\phi}^{Sym}=10^{-20}$, $m_{H^+}=400\,GeV$,
$m_{H^0}=200\,GeV,\, m_{h^0}=100\,GeV,\,m_{A^0}=200\,GeV$,
$sin\alpha=0.1$.} \label{ACPTkphi0}
\end{figure}

Fig. \ref{ACPTkphi0} shows the $Re[k^0_{\phi}]$ dependence of the
$A_{CPT}$ for $k_{\phi\phi}^{Sym}=10^{-20}$, $m_{H^+}=400\,GeV$,
$m_{H^0}=200\,GeV,\, m_{h^0}=100\,GeV,\,m_{A^0}=200\,GeV$,
$sin\alpha=0.1$. Here the $A_{CPT}$ is the order of $10^{-20}$ for
all three different decays since the part without the Lorentz
violating effects in the denominator highly suppresses the ratio
and the behaviors of $Re[k^0_{\phi}]$ for different outputs can
not be distinguished.

At this stage we would like to summarize our results
\begin{itemize}
\item  The Lorentz violation for the decays under consideration is
regulated by CPT even $k^{Sym}_{\phi \phi}$ and CPT odd $k_{\phi}$
coefficients. The latter one is responsible for the tiny CPT
asymmetry.

\item The Lorentz violating part of the $\Gamma$ is more sensitive
to the CPT even coefficient $k_{\phi\phi}^{Sym}$ compared to the
CPT odd one $Re[k^0_{\phi}]$ and it lies in the range $10^{-21}\,
GeV\leq \Gamma \leq 10^{-17}\, GeV$ for $Re[k^0_{\phi}]=10^{-20}\,
GeV$, $10^{-22}\leq k_{\phi\phi}^{Sym} \leq 10^{-18}$ in the case
of $h^0$ output.

\item  There exist a tiny $A_{CPT}$ in the order of $10^{-20}$ for
all three different decays.
\end{itemize}

As a final result, the Lorentz violating effects for the charged
Higgs decays under consideration are too small to be detected in
the present experiments since they depend on the tiny coefficients
which arise from a more fundamental theory at the Planck scale.
\section{Acknowledgement}
This work has been supported by the Turkish Academy of Sciences in
the framework of the Young Scientist Award Program.
(EOI-TUBA-GEBIP/2001-1-8)
\newpage

\end{document}